\documentclass[aps,prl,twocolumn,groupedaddress,amssymb]{revtex4}

\newenvironment{Fig}[1]{
\begin{figure}
\noindent\begin{minipage}[t]{\linewidth}
\begin{center}
\leavevmode
\epsfxsize=\linewidth
\epsfbox{#1}
}
{
\end{center}
\end{minipage}
\end{figure}
}

\begin{document}
\bibliographystyle{prsty}
\input epsf
\title{Influence of the Potential Energy Landscape on the Equilibration and
Specific Heat of Glass Forming Liquids}
\author{Clare C. Yu and Herv\'e M. Carruzzo$^{\dagger}$}
\affiliation{
Department of Physics and Astronomy, University of California,
Irvine, Irvine, California 92697}
\date{\today}
\begin{abstract}
We show that a glass transition, signaled by a peak in
the specific heat vs. temperature, can occur because a glassy 
system that shows no signs of aging progresses so slowly through the
energy landscape that the time needed to obtain an accurate estimate of the
thermodynamic averages exceeds the observation time.
We find that below the glass transition temperature of a 
three dimensional binary mixture of soft spheres, 
the specific heat increases with measurement time spans orders
of magnitude longer than previously recognized equilibration times.
\end{abstract}

\maketitle

As a glass forming liquid is cooled or compressed,
a glass transition occurs when the system falls out
of equilibrium, i.e., when the time scale for
reaching equilibrium exceeds the observation time \cite{Ediger96}.
Like many complex systems, such as proteins and neural networks,
the dynamics of such a system is 
strongly influenced by the potential energy landscape 
where each point corresponds to a particular configuration and energy of 
the system \cite{Goldstein69,Sastry98}. 
The energy landscape can be used to describe the 
three ways in which a system can fall out of equilibrium. 
First a system
can become trapped in a metastable minimum where it stays for the
duration of the observation time. Second a system can be in an 
energetically unlikely part of phase space and proceed slowly to a 
region of the energy landscape where its configurations obey a Boltzmann
distribution. As such a nonequilibrium system evolves toward more probable
regions of phase space, it exhibits aging which means its
properties systematically change with time and do not obey stationarity
\cite{Kob97b}. The aging time, after which aging stops,
is equal to the $\alpha$
relaxation time which is the charactistic time for the system 
to forget its initial configuration. 

The third way is not widely appreciated and is the subject of 
this paper. Namely, even after a glassy system no longer ages 
and has reached basins with appropriate energies,
the system proceeds so slowly through the energy landscape that it
takes a long time to accumulate the large number of statistically independent 
measurements needed to accurately determine a thermodynamic
average. We find from molecular dynamics simulations  
that a glass forming liquid can undergo a
glass transition, as signaled by a peak in the specific heat $C_V$ versus
temperature, that is due to insufficient averaging by a system that
shows no signs of aging. 
The distribution of energies that a system samples in a basin of the energy 
landscape is a subset of the full distribution of energies available to
the system. Since this subset has a smaller variance than the full 
distribution, the resulting specific heat, which is proportional
to the variance of the energy, will be smaller when calculated
from short time spans than from long time spans. These smaller values
account for the values below the peak in $C_V$ on the low temperature
side. Going to longer time spans eliminates the peak, though at temperatures
below the peak temperature $T_p$, these longer time spans can 
be orders of magnitude longer than previously recognized equilibration
times such as the $\alpha$ relaxation time, the 
energy correlation time, and the aging time.

We have performed a molecular dynamics simulation on a 
three dimensional glass forming liquid \cite{Weber85,Kob94} consisting 
of a 50:50 mixture of two types of soft spheres, 
labelled A and B, which differ only in their sizes. 
The interaction between two particles a distance $r$ apart is given by 
$V_{\alpha\beta}(r)
=\epsilon[(\sigma_{\alpha\beta }/r)^{12}+X_{\alpha\beta}(r)]$
where the interaction length
$\sigma_{\alpha\beta }=(\sigma_{\alpha}+\sigma_{\beta})/2$, and the
ratio of the ratio of the diameters
$\sigma_B/\sigma_A=1.4$ ($\alpha$, $\beta =$ A, B). 
The cutoff function $X_{\alpha\beta}(r)=r/\sigma_{\alpha\beta}-\lambda$
with $\lambda =13/12^{12/13}$. The interaction is cutoff at
the minimum of the potential $V_{\alpha\beta}(r)$. 
Energy and length are measured in units of 
$\epsilon$ and $\sigma_A$, respectively. Temperature is in units of
$\epsilon /k_B$, and time is in units of $\sigma_A\sqrt{m/\epsilon}$ where
$m$, the mass of the particles, is set to unity. 
The equations of motion were integrated using the leapfrog
method \cite{Rapaport95} with a time step of 0.005.
During each run the density $\rho_o=N/L^3$ was kept constant.
$N=N_A+N_B$ is the total number of particles. 
The system occupies a cube with volume $L^3$ and periodic boundary conditions.
According to the ideal mode coupling theory, the relaxation time diverges
at a temperature $T_C$ \cite{Gotze92}. 
For our system $T_C=0.303$ \cite{Carruzzo01}.

We cool the system from a high
temperature (T=1.5) by lowering the temperature in steps of $\Delta
T=0.05$. At each temperature we equilibrate for $10^4$ 
time steps and then measure the quantities of interest
for $10^7$ additional steps. 
The results are then averaged over different runs. 
At the glass transition the system falls out 
of equilibrium and becomes trapped in a basin of the energy landscape.
In order to avoid this, we have used parallel tempering together
with molecular dynamics. We implement parallel tempering (PT) 
\cite{Kob01,Carruzzo01} by running
molecular dynamics simulations in parallel at chosen temperatures
using a temperature constraint algorithm \cite{Rapaport95} to keep the 
temperature of each simulation constant. 
At 100 time step intervals we attempt to switch the configurations
of two neighboring temperatures using a Metropolis test which ensures
that the energies of the configurations sampled at any given 
temperature have a Boltzmann distribution. Let $\beta_1$ and $\beta_2$
be two neigboring inverse temperatures, and let $U_1$ and $U_2$ be
the corresponding potential energies of the configurations at these
temperatures. 
If $\Delta=(\beta_1 - \beta_2)(U_2 - U_1)$, then the switch is accepted
with probability unity if $\Delta\leq 0$ and with probability 
$\exp(-\Delta)$ if $\Delta >0$. The acceptance ratio is between
30\% and 75\%. Near $T_p$, the acceptance ratio was above 60\%.
After a swap is accepted, the velocities
of the particles in each configuration are rescaled to suit their 
new temperature. 
Switching configurations allows a given
simulation to do a random walk in temperature space 
in which it visits both low temperatures and high temperatures. This 
helps to prevent it from becoming trapped in a valley of the energy
landscape at low temperatures. Typically we equilibrate for 
2 million time steps and then do measurements for an additional
4 to 10 million molecular dynamics (md) steps. We average over 
different runs which have different initial positions and velocities 
of the particles at each temperature.

We calculate the specific heat $C_V$ per particle at constant volume 
in two ways. The first uses
the fluctuations in the potential energy $U$ per particle: 
$C_V=(3k_B/2) + Nk_B\beta^2\left(\langle U^2\rangle - \langle U\rangle^2\right)$
where the first term is the kinetic energy. The second way uses 
$C_V=d<U>/dT\approx (<U(T_{i+1})>-<U(T_{i})>)/(T_{i+1}-T_{i})$.
The results are shown in Figure \ref{fig:specheat}.
Notice that there is a sharp asymmetric
peak centered at $T_p=0.305\pm 0.003$. The low temperature
side of the peak drops steeply. The curves for 216 and 512 particles
coincide, indicating that there is no size dependence. 
The discrete points are calculated from the energy fluctuations.
The solid line is calculated from the derivative of the energy. The
fact that the two coincide indicates that the system was equilibrated in
all the basins of the energy landscape that were visited.
For comparison we also show the result of cooling through the transition
($\diamond$). The peak in $C_V$ found by cooling
coincides with the peak found in parallel
tempering.
\begin{Fig}{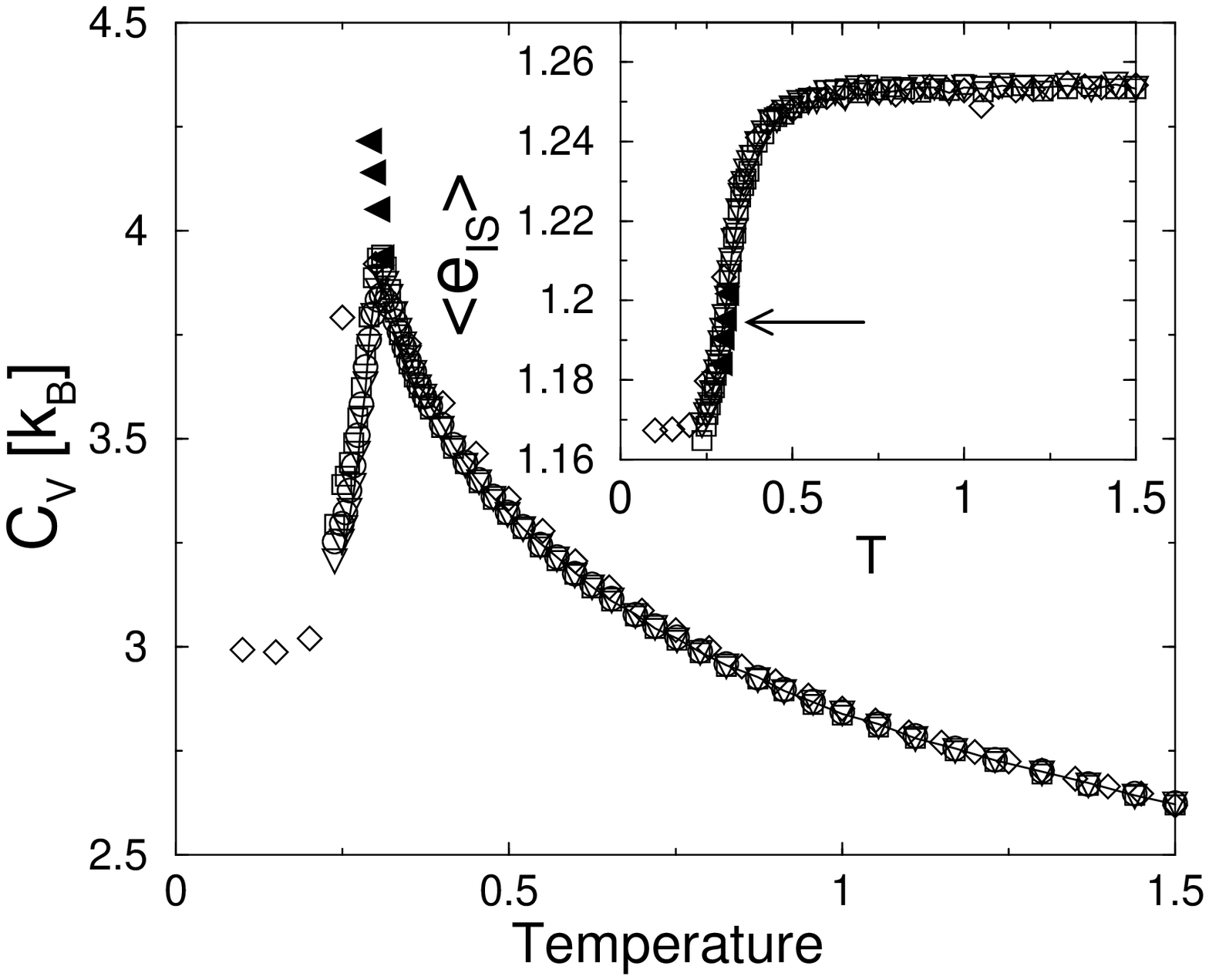}
\caption{Specific heat versus temperature. Shown is
the specific heat calculated from fluctuations for systems
with 216 particles ($\bigcirc$, averaged over 6 runs) and with
512 particles for measurements covering $4 \times 10^6$ time 
steps ($\bigtriangledown$, averaged over 9 runs) and 
$10^7$ time steps ($\square$, averaged over 3 runs) done with
parallel tempering. The solid
line is the specific heat calculated from the
derivative of the energy from the 4 million time step
parallel tempering runs with 512 particles.
Also shown is the specific heat calculated from the
energy fluctuations for a system with 512 particles ($\diamond$,
averaged over 6 runs) that was cooled
conventionally with $10^7$ time steps per temperature.
The $\blacktriangleleft$ correspond to concatenating between 13 and 40
single temperature runs; each run had $10^8$ time steps with
512 particles that were initiated from parallel tempering configurations and
equilibrated for up to $5\times 10^6$ time steps.
Inset: Average inherent structure energy per particle
versus temperature for 512 particles obtained from
parallel tempering, cooling, and single temperature runs. The symbols denote
the same cases as in the main figure. The arrow points to the inflection
point which coincides with $T_p$.
}
\label{fig:specheat}
\end{Fig}
 
We do not believe that the parallel tempering peak in the specific 
heat is due to the system becoming trapped in a metastable minimum in 
the energy landscape for the following reason. 
The configurations corresponding to the bottom of the basins of the
energy landscape are called inherent structures 
\cite{Stillinger82}.
We sampled the configurations that were visited during the 
parallel tempering runs and found the corresponding inherent 
structure energy $e_{IS}$ per particle 
by minimizing the potential energy locally using 
the method of conjugate gradients \cite{Press92}. In 
the inset of Figure \ref{fig:specheat}
we plot the average inherent structure energies versus the temperature
of the configuration that was originally saved.
As the temperature decreases below 0.5, 
$\langle e_{IS}\rangle$ decreases rather steeply \cite{Sastry98}.
The temperature of the inflection point of this decrease coincides with
the temperature $T_p$ of the peak in the specific heat. We also
show $\langle e_{IS}\rangle$ for a system of 512 particles
that was cooled 
from $T=1.5$.
At low temperatures $\langle e_{IS}\rangle$ is rather independent of
temperature for the cooled system, indicating that the system is trapped
in an energy basin. Such a flattening off at low temperatures is not
observed when parallel tempering is used, indicating that 
the system is able to visit deeper energy basins as the temperature
decreases.

However, this does not mean that the peak in the specific heat is
an equilibrium feature. We will now show that the peak is the
result of not sampling enough of phase space below $T_p$.
None of the runs we have done at $T<T_p$ have reached 
the time which is necessary in order to achieve an accurate
thermodynamic average. To show this, we took a
configuration of 512 particles generated by parallel tempering
at $T=0.289855$, equilibrated for $5\times 10^7$ time steps, and 
then ran for an additional $10^8$ time steps during which
we recorded the energy at every time step. Then we did block averaging
in which we divided our 100 million time steps into equal segments,
each of length $\Delta t_b$, and calculated the specific heat from
energy fluctuations for each segment \cite{Ferrenberg91}. The block
averaged specific heat versus time for 2 different time spans $\Delta t_b$ is 
shown in the inset of Figure \ref{fig:sphtimespan}. Notice that
for any given time span, there is no sign of systematic aging.
However, the specific heat averaged over time increases with 
$\Delta t_b$. 

\begin{Fig}{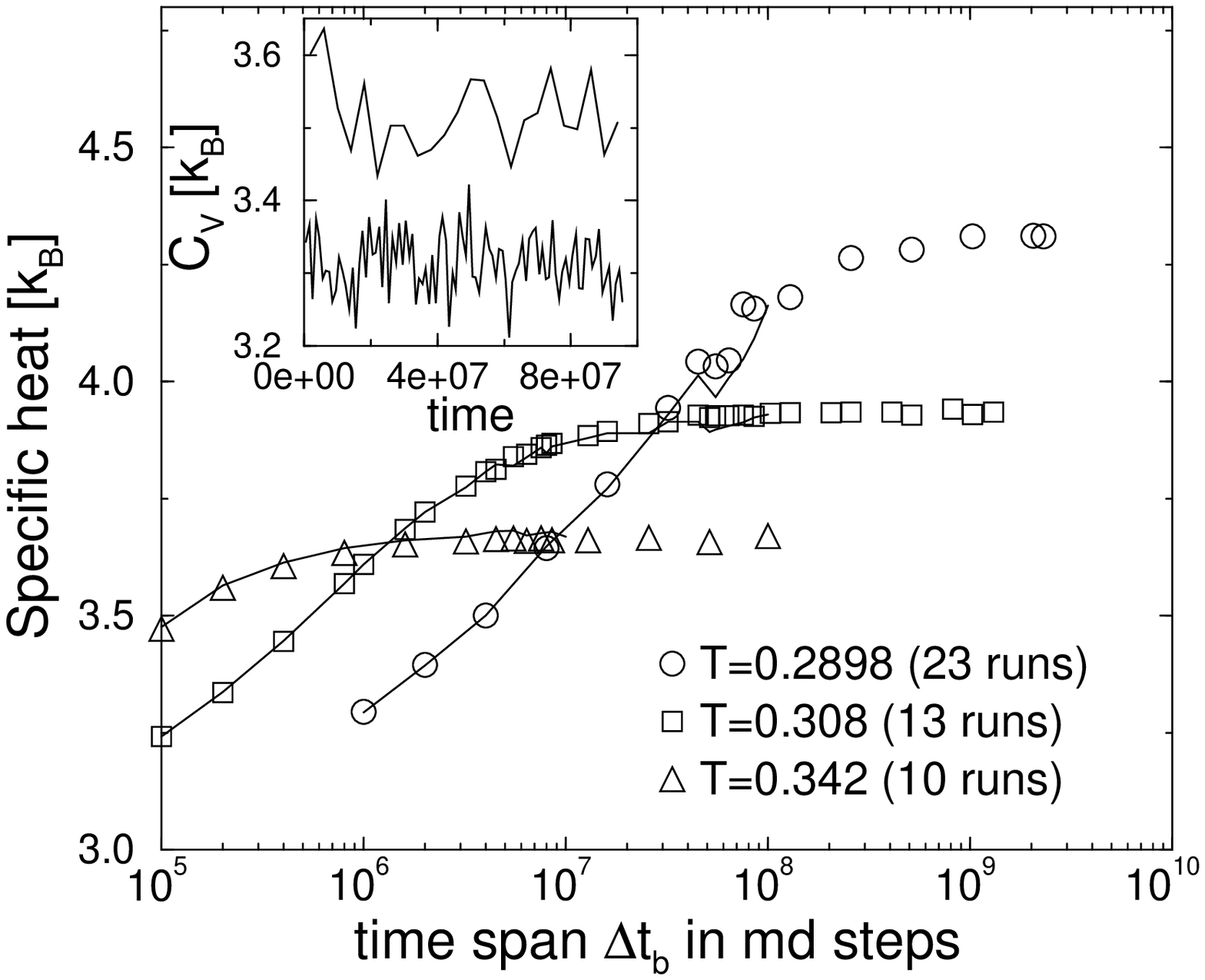}
\caption{Block averaged specific heat versus time span $\Delta t_{b}$
for 512 particles at $T=0.289855$, 0.308642, and 0.3424658.
Solid lines are the result of block averaging each run and
then averaging over the number of runs shown. Open symbols are 
the result of stringing runs together and then block averaging. 
Inset:Block averaged specific heat versus time for
512 particles at $T=0.289855$. 
The lower curve corresponds to $\Delta t_b = 10^6$ md steps, and the upper to
$\Delta t_b=4\times 10^6$ md steps. The time is the time (in md steps)
in the middle of each block. The data at $T=0.289855$ is averaged over 23 runs.
Parameters for both figures are the same as in 
Figure \protect\ref{fig:specheat}.
}
\label{fig:sphtimespan}
\end{Fig}

The specific heat, averaged over time spans of a given size and over different
runs, versus time span size $\Delta t_b$ at several
temperatures is shown in Figure \ref{fig:sphtimespan} by the solid lines.
To obtain time spans that are longer than any given
run, we concatenated the energies from the runs done at a given temperature
to make one huge run, and then did block averaging on the huge run. The
results are shown as open symbols in Figure \ref{fig:sphtimespan}. There
is good agreement between the solid lines and symbols. Notice that the 
specific heat initially increases with time span but then levels off when the
time span is long enough to exceed the 
time needed to achieve the thermodynamic value of the
specific heat. This
time increases with decreasing temperature. Thus at $T=0.289855<T_p$,
the specific heat continues to increase with time span up to
$\Delta t_b = 10^{8}$ time steps which is 100 times longer 
than the $\alpha$ relaxation time $\tau$. Equilibrated values of
$C_V$ near $T_p$ are
plotted in Figure \ref{fig:specheat} and lie above the peak in $C_V$
found with parallel tempering. Thus the specific heat 
peak found with parallel tempering is the result of not sampling
enough of phase space at $T<T_p$ to obtain the true thermodynamic 
value of the specific heat $C^{\rm true}_V$. 
The exploration of the energy distribution
below $T_p$ is slow even with parallel tempering because the probability
of sampling large increases in $U$ are exponentially small.

Note that $C^{\rm true}_V$ is proportional to the variance $\sigma^2_U$ of
the distribution $P(U)$, i.e., $C^{\rm true}_V=N\sigma^2_U/(k_BT^2)$. 
If $\sigma^2_U$ is finite and
if $n$ sample values of $U$ are statistically independent and
identically distributed, then basic statistics dictates that
the measured $C_V$, which is proportional to 
the sample variance $S^2_n$ of $U$, has an expectation value of
$\langle C_V\rangle=C_V^{\rm true}(1-1/n)$ \cite{Ferrenberg91}.
Since the potential energies
in a time series can be correlated, the number $n$ of statistically
independent energies is given by $n=\Delta t_b/\tau_{U}$ where
$\tau_{U}$ is the energy correlation time.
Fitting the data that is within 5 to 10\% of $C^{\rm true}_V$
in Figure \ref{fig:sphtimespan} to 
$\langle C_V\rangle=C_V^{\rm true}(1-1/n)$ yields 
$\tau_{U}\approx 3 \times 10^{6}$, $1 \times 10^5$, and
$5 \times 10^3$ time steps at $T=0.2898551$, 0.308642, and 0.342466,
respectively. These values are comparable to the $\alpha$
relaxation times $\tau$ of $1\times 10^{6}$, $6\times 10^4$, and
$2 \times 10^4$ time steps, respectively (see below). 
They imply that the energies are correlated and that $n$ is substantially 
smaller than the total number of energies.

Note that at shorter time spans and lower 
temperatures Figure \ref{fig:sphtimespan} shows that $C_V\sim \ln(\Delta t_b)$.
This occurs because the system does not uniformly sample $P(U)$
during these shorter time spans. 
As the system travels through the energy landscape, it samples the
energies of each basin that it visits. As shown in 
Figure \ref{fig:distribution}, we find that the distribution of
energies sampled from the basins visited during a shorter time span 
has a smaller variance than the total 
distribution $P(U)$. Furthermore, the centers of the smaller distributions 
do not necessarily coincide with the center of the total distribution.
Rather the smaller distributions are centered at the inherent structure 
energy plus the energy of vibrations around $e_{IS}$ \cite{Schroder00}.
In support of this, we show in the inset of Figure \ref{fig:distribution}
that the block averaged energy versus time approximately
coincides with $(e_{IS}(t)+3k_BT/2)$ vs. time. As more
basins are visited, the sample average $\langle U\rangle$ moves towards the 
average of the full distribution and the variance grows. This corresponds to
$C_V$ increasing with time span. 

\begin{Fig}{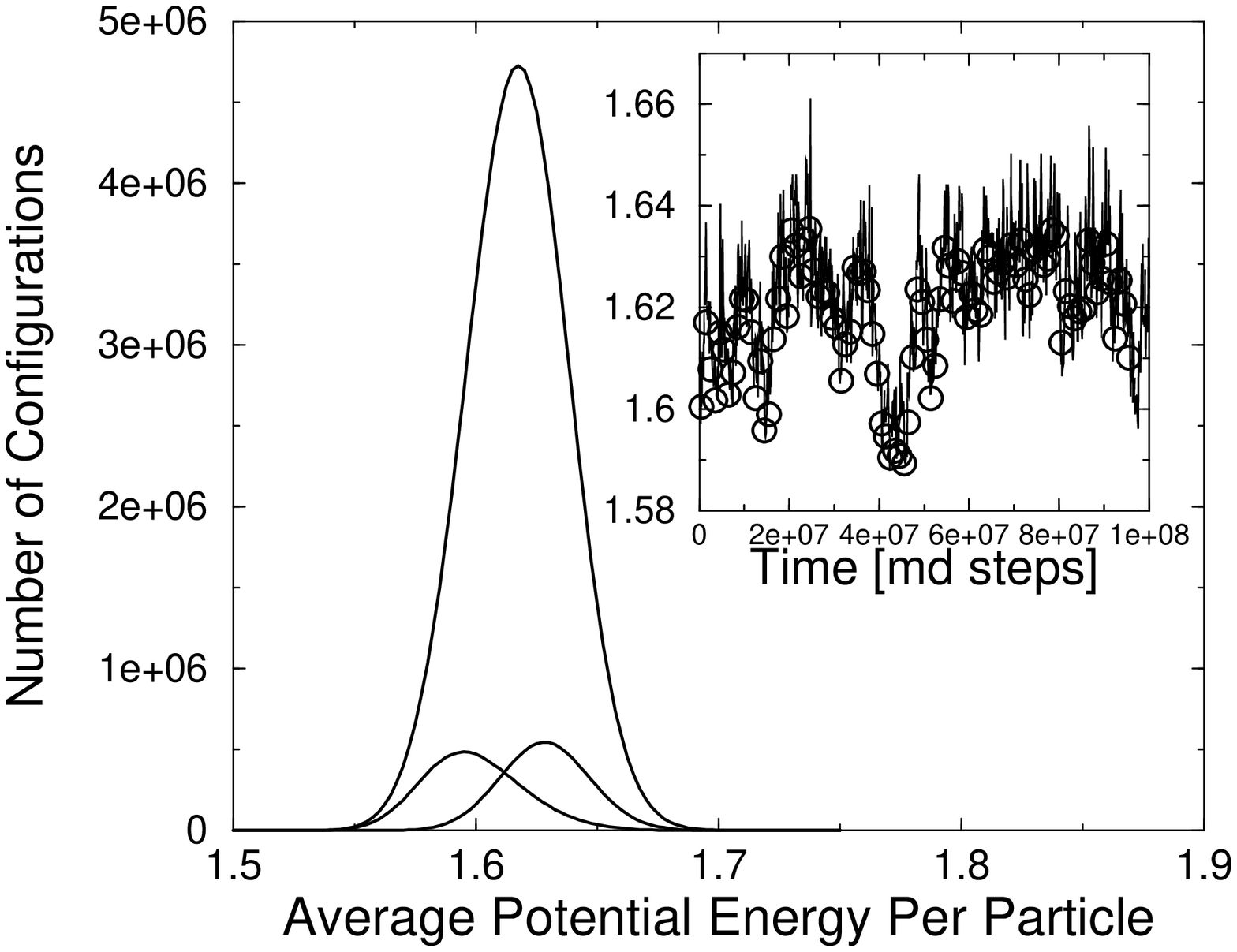}
\caption{Number of configurations versus the average energy per particle
acquired during one run of $10^8$ md steps of a system of 512 particles at 
T=0.289855. The large curve centered at 1.62 is the histogram for the 
entire run. The
small curve on the left centered at 1.60 represents the counts acquired between 
$4\times 10^7$ and $5\times 10^7$ time steps; the small curve on the 
right centered at 1.63 represents the counts acquired between $7\times 10^7$ 
and $8\times 10^7$ time steps.
Inset: Data from the same run. 
Circles represent the potential energy per particle averaged over 
$10^6$ time steps versus time. The solid line is $e_{IS}+3k_BT/2$ versus time. 
}
\label{fig:distribution}
\end{Fig}
 
We now cite evidence that systems at temperatures
just below $T_p$ have equilibrated in the sense of
showing no signs of aging. First the inset of
Fig. \ref{fig:sphtimespan} shows the lack of
aging in the specific heat versus time for a given value of $\Delta t_b$.
Second is the absence of aging in the $\alpha$ relaxation time $\tau$.
We have calculated
$\tau$ using the full intermediate scattering function
$F_{BB}(\vec{k},t)=(1/N_B)\langle\rho_{\vec{k}}(t)\rho_{-\vec{k}}(0)\rangle$
for the B particles where $N_B$ is the number of B particles and
$\rho_{\vec{k}}(t)=\sum_{i=1}^{N_B}\exp(-i\vec{k}\cdot\vec{r}_{i}(t))$.
We choose $k=k_{\rm max}$, the wavevector of the maximum in the
partial static structure factor $S_{BB}(k)$ for the B particles,
because $F_{BB}(k,t)$ relaxes slowest at $k_{\rm max}$ \cite{Rinaldi01}.
The relaxation time $\tau$ is defined by $F(k,\tau)/F(k,t=0)=1/e$.
We have averaged $F(k,t)/F(k,t=0)$ over 40 runs
after waiting times $t_W$ of $5\times 10^7$, 10$^8$, and 
$1.5\times 10^8$ time steps
for a system with 512 particles at $T=0.289855 < T_p$ \cite{Carruzzo01}.
We find that the $\alpha$ relaxation time
$\tau=(1.0 \pm 0.1)\times 10^{6}$ md time steps.
This value of $\tau$ shows no signs of aging \cite{Kob97b} in the
sense that there is no systematic variation with waiting time. 
The lack of aging is to be expected since the aging time is equal 
to $\tau$ which is much less than $t_W$. 
We have confirmed that the aging time is the same as the equilibrium
value of $\tau$ by starting
from 11 different equilibrium configurations at $T=1.5$, quenching
to $T=0.289855$, and measuring $\tau$ after waiting times of
$t_W=0$, $5\times 10^4$, $5\times 10^5$, $10^6$, $5\times 10^6$, and
$2.5\times 10^7$ md steps. For $t_W < 10^6$ md steps, $\tau$ increases
with $t_W$ which indicates aging \cite{Kob97b}. However, for 
$t_W\ge 10^6$ md steps, there is no aging and $\tau$ equals its equilibrium
value of $10^6$ md steps. 

Third we have looked for signs of aging in the inherent structure energy
versus time in our single temperature molecular dynamics runs
of $10^8$ time steps at $T=0.289855 < T_p$. During the run,
configurations were recorded every so
often. Averaging over 40 runs, we find no evidence that the inherent
structure energy decreases systematically with time, though the noise
in the data prevents us from seeing changes smaller than 1\%. 

We have also examined the root mean square displacement
$\langle\Delta r^2(t)\rangle^{1/2}$ where
$\langle\Delta r^2(t)\rangle=(1/N)
\langle \sum_{i=1}^{N}({\bf r}_i(t)-{\bf r}_i(0))^2\rangle$. We found
that $\langle\Delta r^2(t)\rangle^{1/2}\stackrel{>}{\sim} 10\sigma_A$ 
in each run examined after $10^8$ time steps at $T=0.289855$.
This is comparable to the box size L = 9.48$\sigma_A$ for a system
of 512 particles. So the system does not appear to be getting stuck
in a metastable minimum of the energy landscape.

Our work is a cautionary tale for
those who perform numerical simulations on slowly relaxing systems.
It indicates that
to obtain accurate thermodynamic averages, one must not only check
that the system shows no signs
of aging, but one must also check that the quantity to be measured has
sampled enough of phase space to obtain a large number of statistically
independent values. This sampling time can be orders of magnitude
longer than previously recognized time scales such as the aging time
and the $\alpha$ relaxation time.

We thank Francesco Sciortino, Bulbul Chakraborty, Jon Wellner and 
Sue Coppersmith for helpful discussions. This work was supported in 
part by CULAR funds provided by the University of California for the 
conduct of discretionary research by Los Alamos National Laboratory and 
by DOE grant DE-FG03-00ER45843.

$^{\dagger}$Present address: Internap, Atlanta, GA 30309.

\end{document}